# Device JNEEG to convert Jetson Nano to brain-Computer interfaces. Short report

Ildar Rakhmatulin, PhD, electronic researcher
ildarr2016@gmail.com

**Abstract**
Artificial intelligence has made significant advances in recent years and this has had an impact on the field of neuroscience. As a result, different architectures have been implemented to extract features from EEG signals in real time. However, the use of such architectures requires a lot of computing power. As a result, EEG devices typically act only as transmitters of EEG data, with the actual data processing taking place in a third-party device. That's expensive and not compact. In this paper, we present a shield that allows a single-board computer, the Jetson Nano from Nvidia, to be converted into a brain-computer interface and, most importantly, the Jetson Nano's capabilities allow machine learning tools to be used directly on the data collection device. Here we present the test results of the developed device.
https://github.com/HackerBCI/EEG-with-JetsonNano

**Keywords:** JetsonNano, JNEEG, BCI, EEG, machine learning, Nvidia

**Introduction**
Electroencephalography (EEG) is the most popular method of studying the brain. Electrodes placed on the head are used to record data, reading signals in microvolts, which are then digitized and used to extract useful information. This information can be used for a variety of purposes: searching for diseases, monitoring sleep, predicting speech, controlling robotic devices and many others [1,2].
In recent years, a number of companies have set out to develop low-cost devices for reading EEG signals [3,4,5] using non-invasive electrodes [6]. Even then, most of the devices cost more than $1,000. In this paper we present a low-cost device in open-source format that allows the use of CNNs directly in the EEG signal reading board.

**Technical details**
The heart of the JNEEG shield is a Texas Instruments ADS1299 analog digital converter connected to the Jetson Nano via SPI. The ADS299 allows the use of internal amplification and signal transmission up to 16 kSPS. Adjustment of access registers in the software presented by us. The signal reading is implemented in C language and presented as a static library, the data reading and visualisation is implemented in Python, which makes it very easy to set up and modify the code according to the user's needs. During operation it is important that the device is connected to a 5V battery and is completely isolated from the network power supply, this is necessary for safety and to avoid network interference. All parts must be connected to the battery only - monitor, keyboard, mouse, Jetson Nano and JNEEG. General view of the developed device presented in Fig.1.

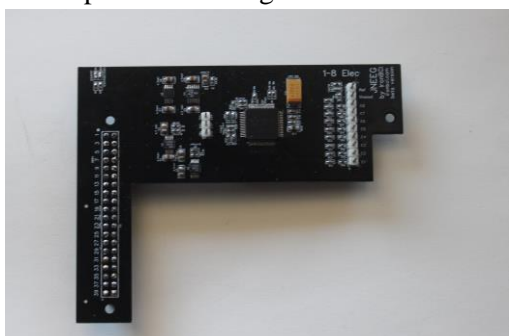

Fig. 1 General view of the JNEEG board

Fig. 2 shows a JNEEG device mounted on a Jetson Nano. The design of the board allows the connection and use of GPIO40 pins of the Jetson Nano and allows the connection of a fan for the Jetson Nano.

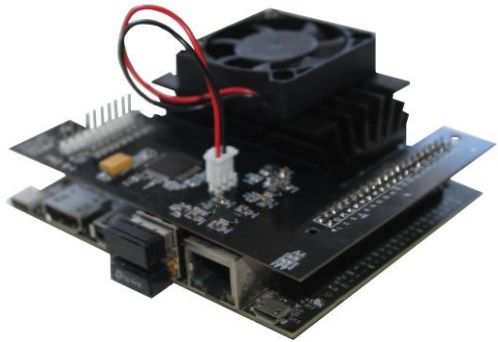

Fig.2. JNEEG board installed on Jetson Nano

The device is capable of measuring 8 channels, but if necessary the design can be modified to allow the addition of «sandwich» format boards, increasing the number of channels read up to 32.

## 3. Experiments

Fig, 3 shows the internal noise test when the pins between the internal ADFS1299 amplifier are short-circuited via the CHnSET register, (address = 05h to 0Ch), which allows you to check the quality of the created PCB design.

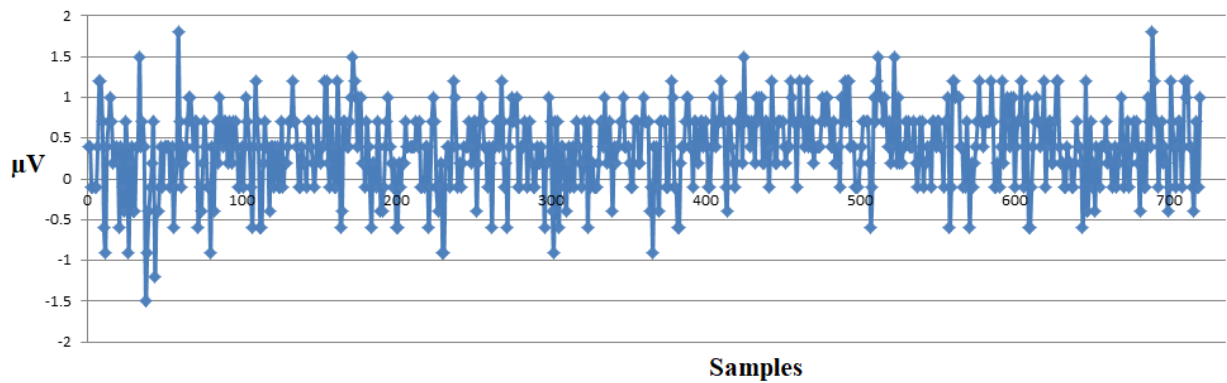

Fig.3. Internal noise test

As a result, the internal noise does not exceed an average of 1 µV.
In Fig. 3, by short-circuiting the pins between the first channel and the reference via electrodes from the outside, it was tested to what extent the device was exposed to electromagnetic interference.

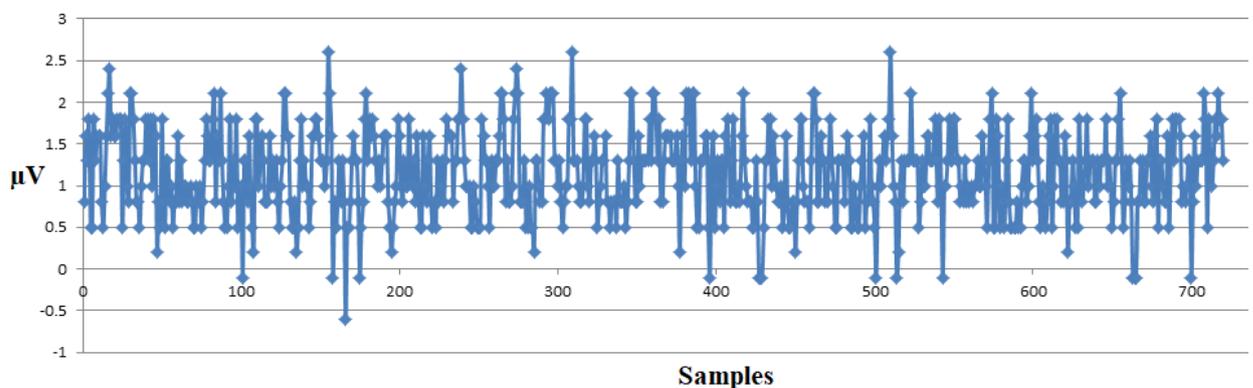

Fig.4. Electrode noise test

The Electrode noise does not exceed an average of 2 µV.

We used 8 electrodes from the international 10-20 system for the following positions - Fz, Cz, Pz, T3, C3, C4, T4, F4. Figure 5 shows the process of measuring the EEG signal for the Fz position after band-pass filtering – 1-40 Hz.

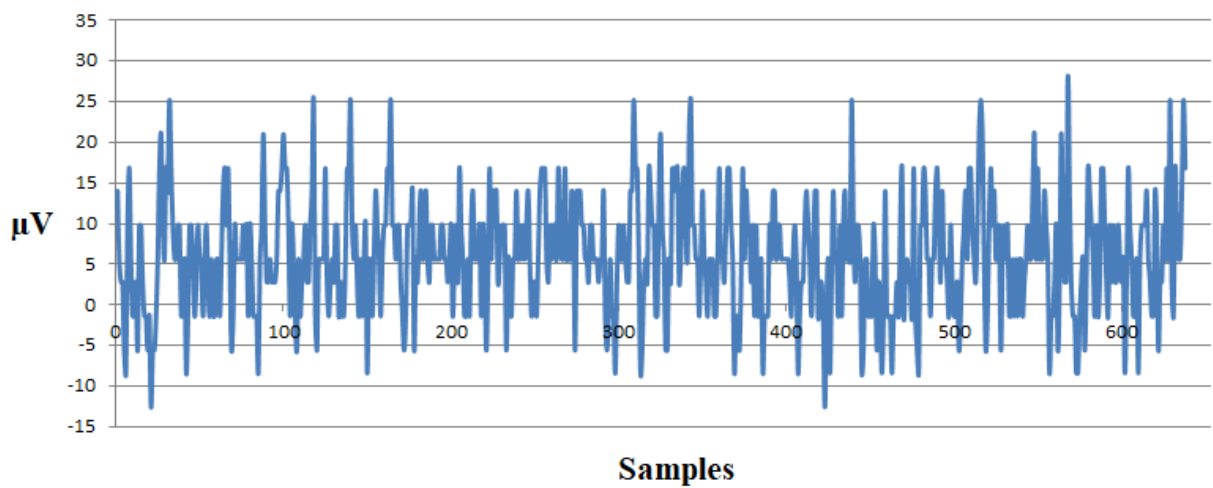

Fig.5. EEG signal from the position Fz

The experiments carried out showed that the proposed device fully complies with the specifications presented for the Texas Instruments ADS1299, confirming the correct design of the implemented JNEEG board.

**Conclusion**
In this article, we have detailed the results of testing the device JNEEG. On our GitHub page we provide detailed instructions and all the necessary technical documentation to allow anyone to assemble this EEG device. We have successfully developed a device with 1.5µV peak input noise and 115dB common mode rejection from 0-50Hz. This device can record an EEG signal with a sampling frequency of 250-1000 Hz. A JNEEG can measure 8 EEG signals simultaneously and the number of electrodes can be increased to 32. A promising future direction of this research is its practical use as a brain-computer interface (BCI). The proposed device, in addition to having a satisfactory noise level and accuracy of detected artifacts in the signal, has also been tested and confirmed that it successfully resists electromagnetic interference and successfully detects alpha brain waves.

**Future plans**
The main advantage of Jetson Nano is the ability to use deep learning, in the following paper we will present the results of research in which we plan to control mechanical objects in real time using the motor imagery method.
]
**CONFLICTS OF INTEREST:** NONE
**FUNDING:** NONE
**ETHICAL APPROVAL:** NOT REQUIRED